\documentclass[
 amsmath,amssymb,
 aps,
prb,
floatfix,
twocolumn, 10pt
]{revtex4-1}

\usepackage{CJKutf8}
\usepackage{graphicx}
\usepackage{bm}
\usepackage{wasysym}
\usepackage{bbold}
\usepackage{xcolor}
\usepackage{hyperref}

\begin{document}

\title{Excitons without effective mass: biased bilayer graphene}

\author{Pengke Li (\begin{CJK*}{UTF8}{gbsn}李鹏科\end{CJK*})}
\email{pengke@umd.edu}
\author{Ian Appelbaum}
\email{appelbaum@physics.umd.edu}
\affiliation{Department of Physics,
U. of Maryland, College Park, MD 20742}

\begin{abstract}
Understanding the dynamics of excitons in two dimensional semiconductors requires a theory that incorporates the essential physics distinct from their three-dimensional counterparts. In addition to the modified dielectric environment, single-particle states with strongly non-parabolic dispersion appear in many two-dimensional band structures, so that ``effective mass'' is ill-defined. Focusing on electrostatically-biased bilayer graphene as an example where quartic (and higher) dispersion terms are necessary, we present  a semi-analytic theory used to investigate the properties of ground and excited excitonic states. This includes determination of relative oscillator strengths and magnetic moments ($g$-factors) which can be directly compared to recent experimental measurements.     
\end{abstract}

\maketitle 

\begin{figure}
\includegraphics[width=2.8in, trim=0 1cm 0 0,clip]{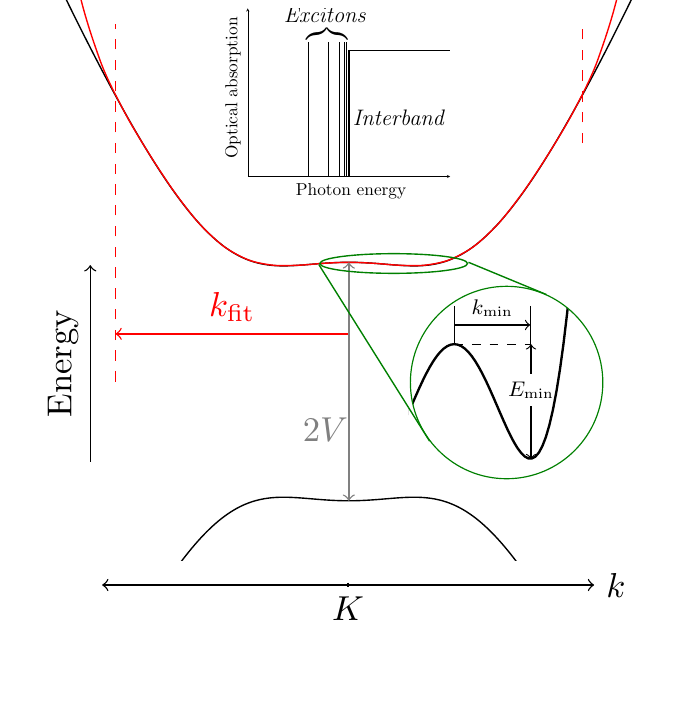}
\caption{Single-particle electronic structure of biased bilayer graphene near the $K(K')$-point under bias $2V=100$~meV shown in black. Red line is least-squares 8th-order polynomial fit within the vertical dashed lines. Magnified region (circled in green) emphasizes the nonparabolic dispersion and energetic depth of band-edge extrema. Inset above shows schematic optical absorption spectrum close to the interband transition threshold, with a manifold of discrete exciton states at lower energy. \label{fig:f1} }
\end{figure}

Analytic solution of the electron Schr\"odinger equation with the attractive Coulomb potential, yielding the Rydberg spectrum of the hydrogen atom, was among the first -- and still monumental --  achievements of quantum mechanics beginning nearly one hundred years ago. Despite its nominal origin in atomic physics, this problem is also very relevant to the solid-state, as a nearly identical mathematical formulation determines the interaction of band electrons and holes with immobile shallow donor and acceptor impurities,\cite{Ramdas_RPP81} and with electrostatic interaction between electrons and holes themselves, resulting in their mutually bound state: \textit{excitons}, somewhat analogous to positronium.\cite{Deutsch_PAAAS53} The presence of these excitons can be indirectly observed in experiments, e.g. optical absorption or photoconduction spectroscopy, as resonances at energy just below the interband excitation edge (see Fig.~\ref{fig:f1}).

The `envelope approximation' often used to model physical attributes of these examples assumes that the effect of absorbing the periodic lattice potential into quasiparticle dispersion only modifies the effective mass, and the lowest-order parabolic relationship between (quasi)momentum and energy remains. However, parabolic dispersion is by no means the only possible outcome endowed by a periodic potential. Especially in two dimensional electronic materials, where weak inter-subband $k\cdot p$ matrix elements suppress otherwise strong band repulsion across a forbidden gap, nonparabolic `Mexican hat' or `caldera'-shaped bands are quite common.\cite{Li_PRB14,Li_PRB15} As shown in Fig.~\ref{fig:f1}, the extrema of these dispersions are indeed approximately quadratic in the radial $k$-direction, but completely flat (ignoring higher order warping from remote bands) in the orthogonal azimuthal direction, yielding a divergent density of states. Such unfamiliar behavior departs considerably from the hydrogen atom problem and cannot be captured by simple mass renormalization.

Motivated by recent experimental measurements of field-tunable exciton spectrum in biased bilayer graphene (BBG),\cite{Ju_Science17} where both electron and hole have nearly identical `caldera' dispersion, we present a general variational theory for the (screened) Coulomb problem in two dimensions when quasiparticle dispersion cannot be captured solely by a single lowest-order parabola $\propto k^2$. This theory allows the calculation of bound state spectrum, oscillator strength, and valley-dependent orbital magnetic moment in a transparent way not dependent on opaque numerical schemes such as density-functional theory (DFT).\cite{Park_NL10}

In general, a two-particle exciton wavefunction can be viewed as the superposition of direct products of electron and hole quasiparticle states in momentum space, weighted by an envelope function. As a result, an exact evaluation of the exciton binding energy through the field-theoretic Bethe-Salpeter equation\cite{Salpeter_PR51} using quasiparticle states from DFT is computationally demanding, and any physical insight into the problem would be obscured behind the numerical details. Our theory focuses on the dominant contributions so that, instead of pursuing absolute precision of the binding energy, it reveals insight into the fundamental exciton physics.  For BBG with an analytic Hamiltonian, our theory is especially important to explain excitonic evolution under electronic structure tuning via external electrical gate bias.

In this theory, the quasiparticle kinetic energy necessarily acquires additional terms (quartic $\propto k^4$ and so on) in higher order, appearing in the effective Hamiltonian through canonical substitution $\bm{k}\rightarrow -i\nabla$\cite{Bhardwaj_IJESRT16,Skinner_PRB16} giving
$H=-A_1\nabla^2 +A_2\nabla^4-A_3\nabla^6+\dots +V(r).$
The coefficients $A_\xi$ of all salient orders can be calculated via least-squares fitting over a test range including the dispersion extrema ($k_\text{fit}$ as shown in Fig.~\ref{fig:f1}), compelled to be self-consistent with the ultimately calculated exciton wavefunction radius in real space.

The presence of nonparabolic terms in the kinetic energy complicates the usual reduction of the two-particle problem to a separable system of relative and center-of-mass coordinates. Furthermore, in the rotationally-invariant caldera dispersion, 
`mass' is not well defined along the azimuthal direction. However, relative position $r=r_e-r_h$ and total momentum $P=p_e+p_h$ are still meaningful quantities. As detailed in Supplemental Material (SM), when both electron and hole have identical dispersions as is nearly the case in BBG, use of canonically conjugate variables $p=(p_e-p_h)/2$ and $R=(r_e+r_h)/2$ allow the two-particle effective (classical) Hamiltonian to be written up to quartic order as  
\begin{align}
&\left[A_1(\frac{1}{2}P^2+2p^2)+
A_2(\frac{1}{8}P^4+3p^2P^2+ 2 p^4)\dots\right]+V(r)\notag\\
&\xrightarrow[]{P=0} 2\left[A_1p^2+
A_2p^4\dots\right]+V(r).
\label{eq:Peq0}
\end{align}

Unlike the usual parabolic kinetic energy case, it is not possible to eliminate all terms that mix momenta $p$ and $P$, so full separation into decoupled equations of motion fails here; in general, the free exciton dispersion will be nonparabolic \textit{and} the exciton wavefunction in relative coordinate $\psi(r)$ will depend on total momentum $P$. 
However, negligible photon momentum requires $P\sim 0$ for analysis of behavior under optical excitation, which is our focus.\cite{Cheianov_PRL12}

When the electron wavefunction is confined to two dimensions, the electrostatic interaction is modified, as initially discovered by Keldysh.\cite{Keldysh_JETP79} There are two asymptotic limits as elaborated by Cudazzo \textit{et al.}\cite{Cudazzo_PRB11}: at large relative distances, the   potential behaves like the usual Coulomb interaction, but close to the origin it diverges only logarithmically. A screening length $r_0$, determined by the 2D polarizability, separates these two limiting behaviors and is an important ingredient in our calculation.
 
By considering photon-induced transition rate and Kramers-Kronig relations in the usual way (see SM), the 2D screening length is generically given by
\begin{align}
r_0=\frac{ q^2\hbar^2}{4\pi^2\epsilon_0 m_0^2}\sum_{\text{c,v}}\int \frac{|P_\text{cv}|^2}{E_\text{cv}^3} d^2k,
\label{eq:r0}
\end{align}
where $q$ is fundamental charge, $\epsilon_0$ is the vacuum permittivity, and $m_0$ is the free electron mass. In addition, the generally $\bm{k}$-dependent terms in the integrand are $P_\text{cv}$ (the matrix element connecting band-edge states of momentum parallel to the electric field) and $E_\text{cv}$ (the gap energy). The denominator of the integrand indicates an inverse relationship between bandgap and 2D polarizability, which further affects the binding energy (in light of the known dependence of the hydrogen spectrum on permittivity).

Our full two-particle Hamiltonian, consisting of nonparabolic kinetic energy operators and the Keldysh form of electron-hole interaction, is not amenable to analytic diagonalization, so a variational method is applied. First of all, in this quasi-rotationally invariant system, the centrifugal term of the Laplacian ($\frac{1}{r^2}\frac{d^2}{d\phi^2}\rightarrow -\frac{m^2}{r^2}$) demands that the wavefunction behave like $r^{|m|}$ for small $r$, where $m$ is the angular momentum quantum number. Using a modified stretched exponential trial function $r^{|m|}\exp[-(r/a)^\beta]\exp(im\theta)$, we find that the expectation value of nonparabolic terms ($\nabla^4$ and higher) requires $\beta\geq 2$ to avoid divergence. Values of $\beta$ significantly greater than 2 would cause a sharp wavefunction suppression for $r>a$ and are therefore unsuitable for trial functions because the asymptotic form far away from the origin (where the potential is approximately Coulombic) should match the Slater-type purely exponential function with $\beta=1$\cite{Yang_PRA91,Prada_PRB15}, except for corrections due to nonparabolicities.

By choosing the $\beta=2$ Gaussian trial envelope wavefunction $\psi_j=r^{|m|}\exp[-(r/b_j)^2]\exp(im\theta)$, we can calculate matrix elements of kinetic energy operators to arbitrary order with
\begin{align}
\langle \psi_i|(-i\nabla)^{2\xi} |\psi_j \rangle= \pi 4^{\xi}(\xi+m)! \frac{(b_i^2b_j^2)^{m+1}}{(b_i^2+b_j^2)^{\xi+m+1}},
\label{eq:gradelems}
\end{align}
where $\xi=0,1,\dots$ indexes powers of the Laplacian. When normalized by the $\xi=0$ inner product, this yields a single-particle variational kinetic energy (for $b_i=b_j=b$) of 
\begin{align}
K_m=\sum\limits_{\xi=1}A_\xi 2^\xi\frac{(\xi+m)!}{m!}b^{-2\xi}.
\label{eq:K}
\end{align}

Evaluating the expectation value of the potential energy requires deeper analysis. Here, we find the integral representation provided by Cudazzo \textit{et al.}\cite{Cudazzo_PRB11} especially useful, where the Keldysh potential is due to a fictitious charge density distributed normal to the plane $q\delta(r)\frac{e^{-|z|/r_0}}{2r_0}$.
As detailed in SM, normalized diagonal matrix elements in the $m=0$ Gaussian basis can be analytically calculated by inverting the order of integration over $r$ and $z$, yielding a generic 2D potential energy
\begin{align}
U_0&= -\frac{ q^2}{8\pi\epsilon_0\bar{\epsilon}r_0} \left[e^{-x^2} \left(\pi  \text{erfi}\left(x\right)-\text{Ei}\left(x^2\right)\right)\right],
\label{eq:U}
\end{align}
where $x^2=\frac{b^2}{8 r_0^2}$ and  $\bar{\epsilon}$ is the relative permittivity of the surrounding medium. 
Here, Ei($x$) is the exponential integral function and erfi($x$) is the imaginary error function. 

An analytic expression for the Keldysh potential matrix element with $m = 1$ is given in SM. For this and higher quantum numbers, the kinetic energy expectation values in Eq.~(\ref{eq:K}) monotonically increase, whereas the potential energy tends to decrease, leading to steadily larger envelope wavefunctions and shallower binding energy.  

Having presented the basic elements of our approach, we now focus on excitons in BBG, whose low energy electronic structure is captured by the four coupled $p_z$ orbitals of both atomic layers, each of which contains two carbon sublattices, $A$ and $B$. We follow the notation of McCann and Koshino,\cite{McCann_RPP13} using the basis ordering $\{A_1,B_1,A_2, B_2\}$ and write the $4\times 4$ tight-binding effective Hamiltonian at the $K$-point as $H_0+H_1+H_2$. 
$H_0$ is the nearest-neighbor $p_z$-orbital Hamiltonian accounting for lowest-order intra-/inter-layer coupling with hopping parameters $\gamma_0=3$~eV and $\gamma_1=0.4$~eV, respectively, and electric-field biasing with on-site energy $\pm V$. This dominant term determines the eigenstates and captures the gross structure of the electron/hole dispersion 
$E = \pm\left[\frac{\gamma_1^2}{2}+\frac{3}{4}(a\gamma_0k)^2+V^2-\frac{1}{2}\sqrt{\gamma_1^4+3(\gamma_1^2+4V^2)(a\gamma_0k)^2}\right]^{\frac{1}{2}}$ (where $a=2.46$~\AA\, is the lattice constant), and is used to extract the polynomial coefficients $A_{\xi}$ used in Eq.~(\ref{eq:K}), within a fitting range of several times $k_\text{min}$ (see Fig.~\ref{fig:f1}). Additional terms $H_1$ and $H_2$ have only a minor effect on the energy dispersion and the eigenstates, but are essential perturbations to include in understanding the exciton oscillator strength and orbital magnetic moment. The former reflects next-nearest-neighbor interlayer ``skew'' coupling $\gamma_3 = 0.3$~eV between non-dimer sites, resulting in trigonal warping of the bands. The remaining term $H_2$ is responsible for the electron-hole dispersion asymmetry, including the dimer/nondimer on-site asymmetry energy $\Delta'\approx 0.02$~eV and the skew interlayer coupling $\gamma_4 = 0.14$~eV between a non-dimer and a dimer site. 
Full matrix expressions for the Hamiltonian are given in SM.

\begin{figure}
\includegraphics[width=3.25in]{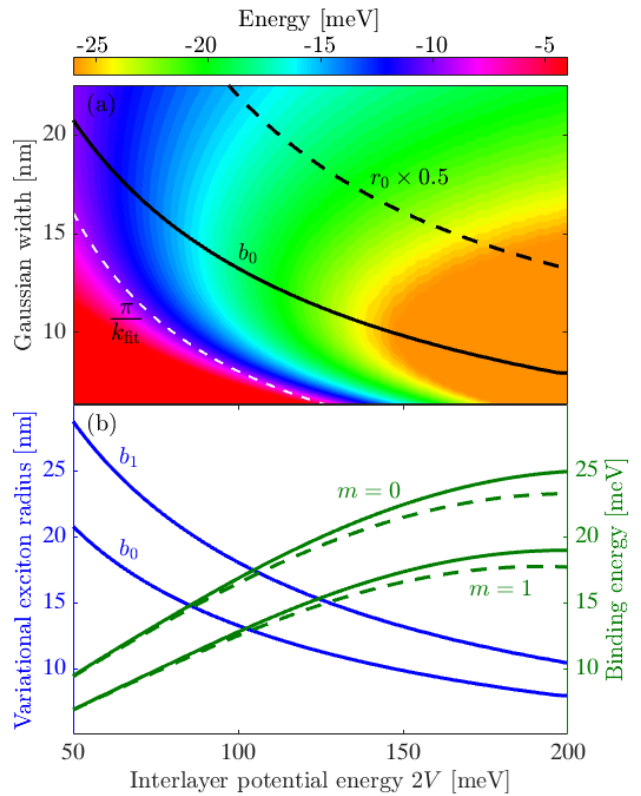}
\caption{\label{fig:f2} (a) Energy landscape of a single $m=0$ Gaussian trial function for the ground-state exciton envelope function of BBG, using Eqs.~(\ref{eq:K}) and (\ref{eq:U}). Kinetic energy polynomial coefficients $A_\xi$, where $\xi=1,2,3,4$, are determined by least-squares fitting the exact dispersion between our choice of $\pm k_\text{fit}$, giving a real-space lengthscale shown as a dashed white line. The energetic minima, indicating variational optimum lengthscale $b_0>\pi/k_\text{fit}$, is shown as a solid black line. Black dashed line presents half of the 2D screening length $r_0>b_0$. (b) Variational optimum lengthscale shown in blue (left axis) for both ground-state ($b_0$) and $m=1$ excited state ($b_1$). Binding energy in green (right axis) accounts for reduction in bound-continuum energy due to $E_\text{min}$ (see Fig. 1). Dashed lines are energies of the single trial function, which are improved by the lowest generalized eigenvalue of the problem with an optimized 5-function basis (see text).  }
\end{figure}

The simplicity of $H_0$ allows analytic evaluation of the momentum matrix element between the conduction and valence bands, and hence the screening length $r_0$ using Eq.~(\ref{eq:r0}). We have (see detailed calculation in SM)
\begin{align}
|\langle P_\text{cv}&(k,\phi)\rangle|^2=\nonumber\\ &\frac{9\gamma_1^2(a\gamma_0 k)^2(V^2\cos^2\phi+E^2\sin^2\phi)}{4[\gamma_1^4+4(\gamma_1^2+4V^2)(a\gamma_0 k)^2]E^2}\left(\frac{m_0a\gamma_0}{\hbar}\right)^2\label{eq:Pcv},
\end{align}
where $\phi$ is the angle between quasimomentum $\bm{k}$ and the photon polarization (chosen as parallel to the $x$-axis). 
Note that in the small $k$ limit, this expression reduces to $\left|\frac{3m_0(a\gamma_0)^2k}{2\hbar\gamma_1}\right|^2$, independent of $\phi$ and consistent with first-order perturbation theory. This result is notable for the absence of optical coupling across the fundamental bandgap at the $K$-point ($k=0$).\cite{Yao_PRB08} It is often the case that symmetry is responsible for vanishing matrix elements, but here no such constraint exists. 
As we will show, symmetry-allowed terms in perturbation $H_1$ are responsible for nonzero interband optical coupling and a bright $m=0$ exciton in real BBG.

With $P_\text{cv}$, the integration for the screening length $r_0$ in Eq.~(\ref{eq:r0}) is straightforward but yields a cumbersome expression (see SM). Graphically, however, it is a featureless curve, as shown by the dashed black line in Fig.~\ref{fig:f2}(a); this lengthscale should be compared to 
the exciton Gaussian width discussed below. Clearly, with increasing gate bias and larger $|V|$, $r_0$ decreases mainly due to the increased band gap.

The transcendental functions in Eq. (\ref{eq:U}) with  $r_0$ as an input require a numerical minimization of the total energy $K_0+U_0$ to find the optimum value of variational parameter $b$. In Fig. \ref{fig:f2}(a) we plot the energy of the $m=0$ Gaussian exciton for $\bar{\epsilon}=4$ (appropriate for BN encapsulation)  as a function of electric field bias, and indicate the lengthscale $b_0$ that minimizes it with a solid curve. The dashed white curve is the equivalent lengthscale determined by the reciprocal of the polynomial fitting region $\pi/k_\text{min}$, showing consistency with our initial assumptions. 

This exciton size variation with bias field is reproduced in Fig.~\ref{fig:f2}(b) as a solid blue line, along with the equivalent result for $m=1$. Both indicate increased confinement with gate bias, consistent with increasing variational binding energies (dashed green lines) of both excitons using a single trial wavefunction. 
To improve upon the single-function variational binding energies, we augment the basis with four additional functions of the same form but with optimized exponentially-spaced lengthscales\cite{Ditchfield_JCP70} and solve for the lowest generalized eigenvalue of $\langle i|H|j\rangle \Psi=E\langle i|j \rangle \Psi $, using Eq.~(\ref{eq:gradelems}) and a generalization of Eq.~(\ref{eq:U}) where $2/b^2\rightarrow(1/b_i^2+1/b_j^2)$. Binding energies calculated in this way (solid green lines) can typically be improved by only less than a few percent, indicating the suitability of the chosen gaussian-type basis for this problem. The magnitude difference of the two exciton binding energies (several meV) and its gate bias dependence have good agreement with the experimentally-measured value.\cite{Ju_Science17}

\begin{figure}
\includegraphics[width=3.25in]{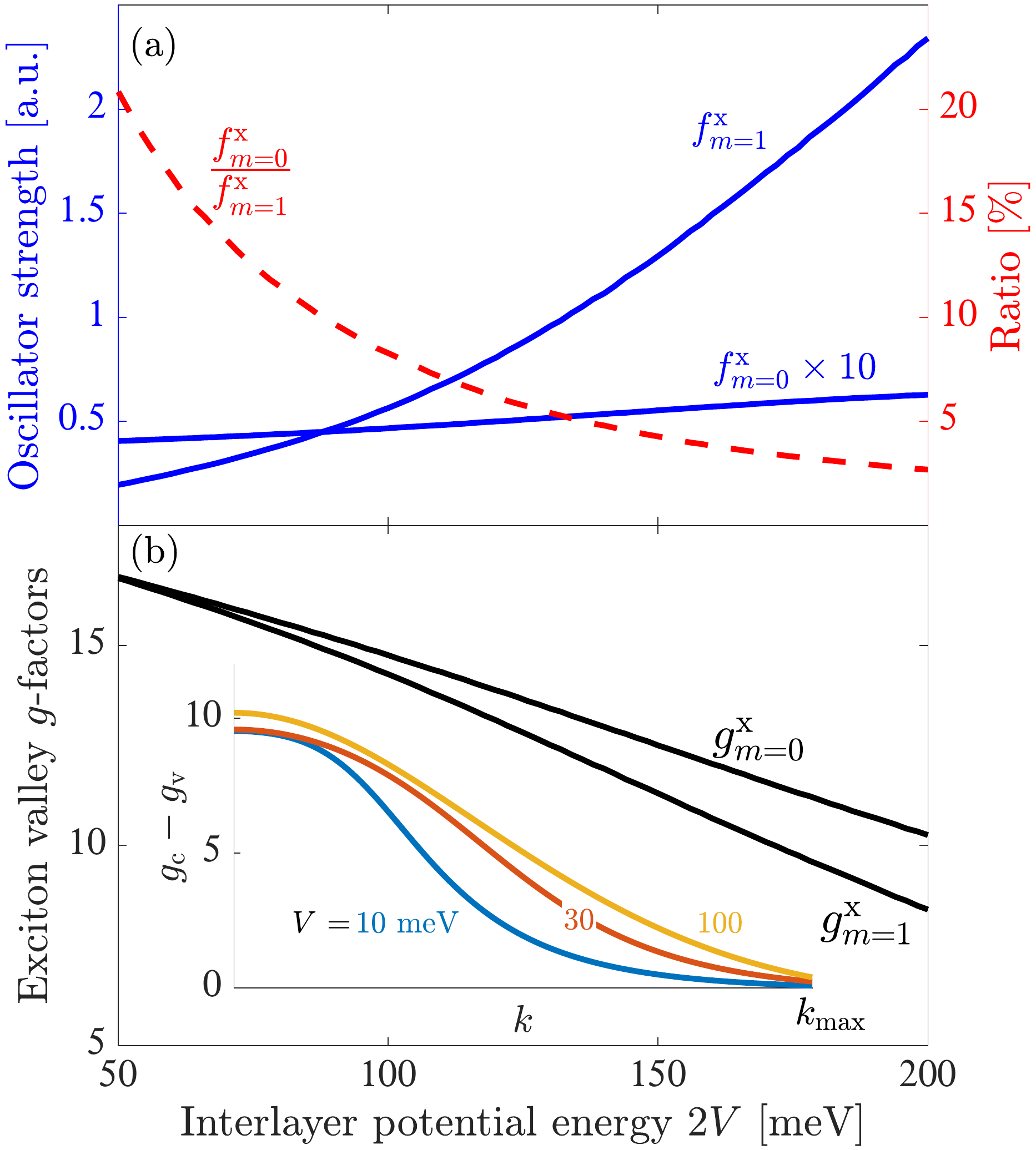}
\caption{(a) Relative oscillator strengths of the $m=0$ and $m=1$ excitons (blue curves) and their ratio (red dashed curve), as a function of the gate bias energy $2V$. (b) Valley $g$-factors of the $m=0$ and $m=1$ excitons. Inset: $k$-dependent $g$-factor differences of the conduction and valence bands, under bias conditions $V=$ 10, 30 and 100 meV. Here, the full scale of $k$ is normalized according to $\frac{\sqrt{3}}{2}a\gamma_0 k_\text{max} = \frac{\gamma_1}{2}$, so that $k_\text{max}$ is less than 2\% from the $K$ point to the $\Gamma$ point.
\label{fig:os_and_gf} }
\end{figure}

Our envelope wavefunctions can now be used to examine the exciton ``brightness'', by evaluating the oscillator strength $f^\text{x}_m\propto |\int \Phi_m(k) P_\text{cv} d^2k|^2/E_\text{x}$, \cite{Yu_B10} where $\Phi_m(k)$ is the Fourier transform of exciton envelope function, and $E_\text{x}$ is the excitation energy of the exciton. Here we employ the L\"owdin partitioning method to reduce the full $4\times 4$ Hamiltonian to a $2\times 2$ matrix in the non-dimer $\{A_1,B_2\}$ basis that captures the two gap-edge bands.\cite{McCann_RPP13} Considering only the dominant term $H_0$, the eigenstates of this two-level system are
\begin{align}
|\text{c}\rangle = \left[
\begin{array}{c}
\cos\frac{\eta}{2}  \\
e^{2i\phi}\sin\frac{\eta}{2}
\end{array}
\right], \,\,\,\,\text{and}\,\,\,\,|
\text{v}\rangle = \left[
\begin{array}{c}
-e^{-2i\phi}\sin\frac{\eta}{2}  \\
\cos\frac{\eta}{2}
\end{array}
\right],
\end{align}
where $\eta\approx 3(a\gamma_0)^2k^2/4V\gamma_1$ (which vanishes at the $K$-point). We must emphasize here that, to maintain the adiabaticity of the wavefunction through the $K$-point, $\bm{k}$-dependent phase factors $e^{\pm2i\phi}$ \textit{should not} be assigned arbitrarily among the components of the states\cite{Park_NL10}, which is crucial in determining the exciton optical selection rules (see SM). In this band basis, the interband matrix element of the momentum operator $\frac{m_0}{\hbar}\nabla_k(H_0+H_1)$ is
\begin{align}
P_\text{cv}
\!\approx\! \frac{m_0}{2\hbar}\left[\sqrt{3}a\gamma_3\!-\!\frac{3a^2\gamma_0^2\gamma_1}{\gamma_1^2+V^2} k e^{-i\phi}\right].
\end{align}
The two bracketed terms play different roles due to their parity. 
Specifically, the first ($k$-independent) and the second ($k$-linear) terms are relevant to the $\Phi_{m=0}$ and $\Phi_{m=1}$ envelope wavefunctions, respectively, to produce nonvanishing  azimuthal integration of $\Phi_mP_\text{cv}$. Importantly, electromagnetic coupling of the $m=0$ exciton ground state depends crucially on the next-nearest-neighbor interlayer coupling parameter $\gamma_3$. Both oscillator strengths increase as a function of gate bias, as shown in Fig.~\ref{fig:os_and_gf}(a). Since $\gamma_3\approx 0.1\gamma_0$, $f^\text{x}_{m=0}$ is one order of magnitude smaller than $f^\text{x}_{m=1}$, even though the single particle excitation of the latter is of higher order in $k$. At large gate bias when both excitons share similar $E_\text{x}$, the ratio of their oscillator strengths can be estimated solely from integration of $\Phi_mP_\text{cv}$ (see SM), 
\begin{align}
\frac{\left(\frac{\sqrt{6\pi} m_0a\gamma_3}{\hbar b_0}\right)^2}{\left(\frac{12\sqrt{\pi} m_0a^2\gamma_0^2\gamma_1}{\hbar b_1^2(\gamma_1^2+V^2)}\right)^2} = \left(\frac{1}{2\sqrt{6}}\frac{b_1^2}{b_0a}\frac{(\gamma_1^2+V^2)\gamma_3}{\gamma_0^2\gamma_1}\right)^2.\label{eq:OS_ratio}
\end{align}
For example, with a gate bias $2V = 100$~meV, using the variational values [see Fig.~\ref{fig:f2}(b)] $b_0 \approx 13$~nm and $b_1\approx 18$~nm, Eq.~(\ref{eq:OS_ratio}) gives a ratio of $\sim 8\%$ that matches well with experimental observation. \cite{Ju_Science17} 

Lastly, we examine the exciton magnetic susceptibility. Similar to positronium,\cite{Deutsch_PAAAS53} the  angular momentum of the envelope function has diminished contribution to the magnetic moment, due to the similar dispersion but opposite charge of the electron and hole. 
On the other hand, the difference between conduction and valence quasi-particle orbital $g$-factors can contribute to the magnetic susceptibility through the Bloch part of the exciton wavefunction. Indeed, electron-hole asymmetry is induced by perturbation $H_2$, resulting in an exciton \textit{valley} $g$-factor due to the opposite magnetic moments at time-reversed $K$ and $K'$ valleys.

The orbital magnetic moment of a quasiparticle state \cite{Dresselhaus_book}
\begin{align}
g_n(\bm{k})\mu_B = i\frac{e\hbar}{2m_0^2}\sum_{\ell\neq n}\frac{\bm{P}_{n\ell}(\bm{k})\times \bm{P}_{\ell n}(\bm{k})}{E_n(\bm{k})-E_\ell(\bm{k})}\label{eq:gfactor}
\end{align}
is identical for the two bands in a generic two level system, so we return to the full $4\times 4$ Hamiltonian and treat $H_2$ perturbatively.
The difference between $g$-factors of the electron and hole states is analytic at the $K$-point, 
\begin{align}
\!\!\!\!g_\text{v}\!-\!g_\text{c} \!\approx\! \frac{3m_0a^2}{\hbar^2\gamma_1^2}\!\left[\gamma_0^2\Delta'\!\left(\!1\!+\!\frac{4V^2}{\gamma_1^2}\!\right)\!+\!2\gamma_1\gamma_0\gamma_4\right]\approx 10, \label{eq:valley_g}
\end{align}
composed of two contributions within square brackets (see SM). The first one $\propto\Delta'$ is due to the dimer-nondimer onsite asymmetry resulting in different energy denominators for conduction and valence bands in Eq.~(\ref{eq:gfactor}). The remaining part is more dominant, involving interference between the $\gamma_0$- and $\gamma_4$-dependent matrix elements in the momentum operator, as evident by their product. As a result, the difference between $g_\text{v}-g_\text{c}$ at $K$ and $K'$ is $\sim 20$.
The $k$-dependent $g$-factor difference of the conduction and valence bands is shown in the inset of Fig.~\ref{fig:os_and_gf}(b) under three different bias fields. As expected, the energy denominators between the gap edge bands and remote bands in Eq.~(\ref{eq:gfactor}) increase as $k^2$ and quickly suppress the value of $g_\text{v}(k)-g_\text{c}(k)$ at large $k$.

The exciton valley $g$-factors contributed by the Bloch wave part are calculated (see SM) by 
\begin{align}
g^\text{x}_m=2\int |\Phi_m(k)|^2 [g_\text{c}(\bm{k})-g_\text{c}(\bm{k})] d^2\bm{k}
\end{align}
for both the $m=0$ and $m=1$ excitons, and presented in Fig.~\ref{fig:os_and_gf}(b) as a function of the gate bias. As $V$ increases, excitons are more confined with smaller radii and larger $k$-space distributions of their envelope wavefunctions, which reduce the valley $g$-factors. $g^\text{x}_{m=1}$ decreases faster than $g^\text{x}_{m=0}$ since $\Psi_{m=1}$ is linear in $k$ and further suppresses the contribution around $k=0$. Note that the Bloch wave contributions to  both exciton $g$-factors do not closely match the experimentally observed large $g$-factor $\sim20$ for $m=0$ and a negligible magnetic susceptibility for $m=1$ excitons.\cite{Ju_Science17} In that experiment, broadband excitation of a relatively high density of excitons and free carriers may push the system into a strong correlation regime, where many-body interaction causes significant deviation from the expected behavior of an isolated exciton. 
This extension to our theory, however, is beyond the scope of discussion in this Letter. 

We end by emphasizing the generality of Eqs.~(\ref{eq:r0}), (\ref{eq:gradelems}), (\ref{eq:K}) and (\ref{eq:U}) applied to excitons in an arbitrary two-dimensional semiconductor with approximately rotationally-invariant nonparabolic bands, such as the valence band in $D_{3h}$ three-six-enes Ga$_{1-x}$In$_x$S$_y$Se$_{1-y}.$\cite{Li_PRB15} Other deviations from parabolic dispersion abound, including Rashba spin-split bands\cite{Rashba_FTT59, Bihlmayer2015, Skinner_PRB16} and anisotropic examples of recent interest such as in the valence band of  phosphorene\cite{Li_PRB14,Prada_PRB15} or the `camel-back' valence band in 3D bulk tellurium,\cite{Li_PRB18} for which our matrix element expressions can be appropriately modified. 

We acknowledge support from the National Science Foundation under contract ECCS-1707415.

\bibliography{exciton}

\end{document}